\documentstyle[aps,epsfig,preprint,prb]{revtex}

\begin{document}


\draft


\title{Scaling of $H_{c2 \perp}(T)$ in Nb/CuMn Multilayers}

\author {A. Angrisani Armenio, L. V. Mercaldo, S. L. Prischepa,\thanks{
Permanent address: State university of Informatics and
RadioElectronics, P. Brovka Str. 6, 220600 Minsk, Belarus} M.
Salvato, C. Attanasio\thanks{Corresponding author.
Telephone:+39-089-965313; Fax: +39-089-953804; E-mail:
attanasio@sa.infn.it}, and L. Maritato}
\address{Dipartimento di Fisica and INFM, Università
degli Studi di Salerno, Baronissi (Sa), I--84081, Italy.}

\date{\today}
\maketitle

\begin{abstract}

Measurements of the perpendicular upper critical magnetic field
$H_{c2 \perp}(T)$ are reported for several Nb/CuMn multilayers.
It is found that, despite the magnetic nature of the samples, the
data for samples with low Mn percentage in the CuMn layers are
simply described by the Werthamer-Helfand-Honenberg theory for
conventional type-II superconductors, neglecting both Pauli spin
paramagnetism and spin orbit impurity scattering. For high Mn
concentration a different theoretical aprroach is needed.
\end{abstract}

\pacs{Key words: superconductivity, spin glass, multilayers,
critical magnetic field. Running head: Nb/CuMn multilayers}

\section{INTRODUCTION}

The issue of the proximity effect in superconducting multilayers
has been intensely studied since the early
sixties\cite{Degennes,Hauser1}. In particular since
superconductivity and magnetism are two mutually excluding
phenomena a lot of interest has been devoted to the analysis of
superconducting (S)/magnetic (M) multilayers\cite{Hauser2,Jin}.
Several theoretical and experimental studies have been done on
these systems, in particular about the existence of the so-called
$\pi-phase$ state which manifests itself in a nonmonotonic
behavior of the transition temperature $T_c$ of the S/M
multilayers as a function of the magnetic layers thickness
$d_M$\cite{Radovic1}. Numerous experiments about the behavior of
$T_c$ versus $d_M$ have been reported on different S/M
multilayers\cite{Korevaar,Verbank,Strunk,Jiang,Mughe,Mercaldo,Attanasio}:
However the presence of $T_c$ oscillations is still an open
question and further theoretical investigation is needed. In
particular in the case of Nb/CuMn multilayers (where CuMn is a
spin glass) the presence of a small Mn percentage ($\geq$ 0.7 \%)
in copper gives rise to a nonmonotonic behavior of $T_c$ versus
$d_M$ \cite{Mercaldo,Attanasio} which cannot be explained in the
framework of the conventional proximity theory even when taking
into account a paramagnetic pair breaking mechamism\cite{Hauser2}.

Another interesting feature of S/M multilayers is the temperature
behavior of the upper critical magnetic field, both in the
direction parallel, $H_{c2 \parallel}(T)$, and perpendicular,
$H_{c2 \perp}(T)$, to the plane of the film, which shows
deviations from the S/N case (here N is a normal metal).
Measurements performed on V/Fe multilayers\cite{Korevaar}
revealed a good agreement with the theoretical predictions for
S/M multilayers\cite{Radovic2}. Both $H_{c2 \parallel}(T)$ and
$H_{c2 \perp}(T)$ could be consistently described using the same
value for the only free parameter of the theory. On the other
hand the same measurements performed on Nb/CuMn samples with 2.7
\% and 4.5 \% of Mn could be only qualitatively described by the
same theory, probably due to the much more complicated nature of
a spin glass system with respect to the ferromagnetic
case\cite{PhysicaC1}.

In this paper we report on measurements of the perpendicular
upper critical magnetic field as a function of temperature in
Nb/CuMn multilayers. A large number of samples have been measured
with different Mn percentage and different layering, also in the
presence of a regular array of
antidots\cite{PhysicaC1,Jap,Philmag,Prbsubmitted}. Regardeless of
the specific nature of the multilayers, the measurements for the
samples with low percentage of Mn in the CuMn layers (up to 2.7
\%) are in agreement with the Werthamer-Helfand-Honenberg (WHH)
theory, which describes the behavior of conventional type-II
superconductors\cite{WHH}. This result indicates that $H_{c2
\perp}(T)$ measurements are less sensitive to the presence of Mn
than the measurements of $T_c$ versus the magnetic layers
thickness. However for sufficiently high Mn concentration a
different theoretical approach is needed to describe the data.

\section{EXPERIMENT}

Nb/CuMn multilayers have been fabricated by using a dual-source
magnetically enhanced dc triode sputtering system with a movable
substrate holder on Silicon (100) substrates\cite{Mercaldo}. The
bottom layer is CuMn and the top layer is Nb for all the samples.
Some of the samples are patterned into 200$\times$200 $\mu$m$^2$
zones with a regular array of antidots and suitable contact pads.
The preparation details for these samples, obtained by lift-off
procedure, are reported elsewhere\cite{Philmag,Prbsubmitted}. All
the samples present good superconducting properties and a well
defined layered structure as shown by low angle X-ray diffraction
patterns\cite{PhysicaC1}.

\begin{figure}[h]
\vskip2cm
\begin{center}
\leavevmode \epsfxsize=15cm \epsfysize=10cm \epsffile{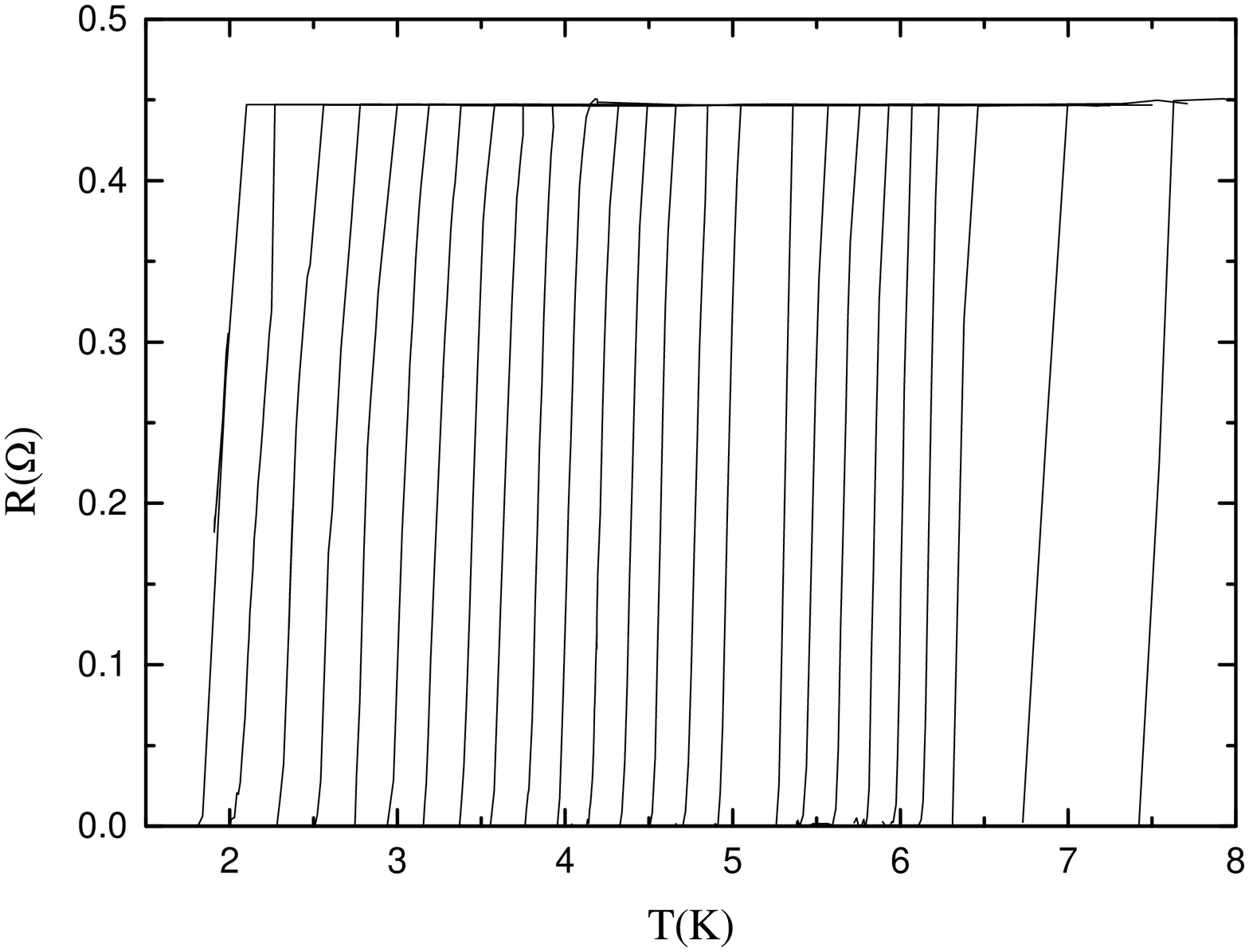}
\end{center}
\caption{\label{fig1}{Transition curves for different
perpendicular magnetic fields for the sample A(20)1. The curves
correspond to increasing fields, from right to left, equal to
0.0, 0.4, 0.7, 0.8, 0.9, 1.0, 1.1, 1.2, 1.5, 1.6, 1.7, 1.8, 1.9,
2.0, 2.1, 2.2, 2.3, 2.4, 2.5, 2.6, 2.7, 2.8, 2.9, 3.0 T.}}
\end{figure}

Transport measurements have been performed with a standard {\sl
dc} four probe technique with magnetic field applied
perpendicular to the plane of the film. In figure \ref{fig1} the
resistive transitions of one of the analyzed samples (sample
A(20)1) are reported at different values of the external
perpendicular magnetic field. The $H_{c2 \perp}(T)$ values have
been obtained from the $R(T)$ curves, at different applied
magnetic fields, using the 50\% $R_N$ criterion, where $R_N$ is
the normal state resistance just before the transition to the
superconducting state. However, even if different criteria are
used to extract the $H_{c2 \perp}(T)$ from $R(T,H)$ curves, no
substantial differences are observed in the results. We have also
occasionally performed $R(H)$ measurements at different
temperatures to extract the $H_{c2 \perp}$ value for each
temperature at the intersection point between the flux flow
regime and the normal state resistance\cite{Larkin}. The critical
magnetic field values obtained with the two different methods are
always in good agreement with each other.

Table I shows the sample characteristics: Nb thickness $d_{S}$,
CuMn (Cu) thickness $d_M$, Mn percentage, number of bilayers,
superconducting critical temperature $T_c(K)$ and anisotropy
ratio $\zeta = H_{c2 \perp}(0)/H_{c2 \parallel}(0)$. A column is
added in the end to point out the patterned samples with the
array of antidots.

\section{RESULTS AND DISCUSSION}

\begin{figure}[h]
\begin{center}
\leavevmode \epsfxsize=15cm \epsfysize=10cm \epsffile{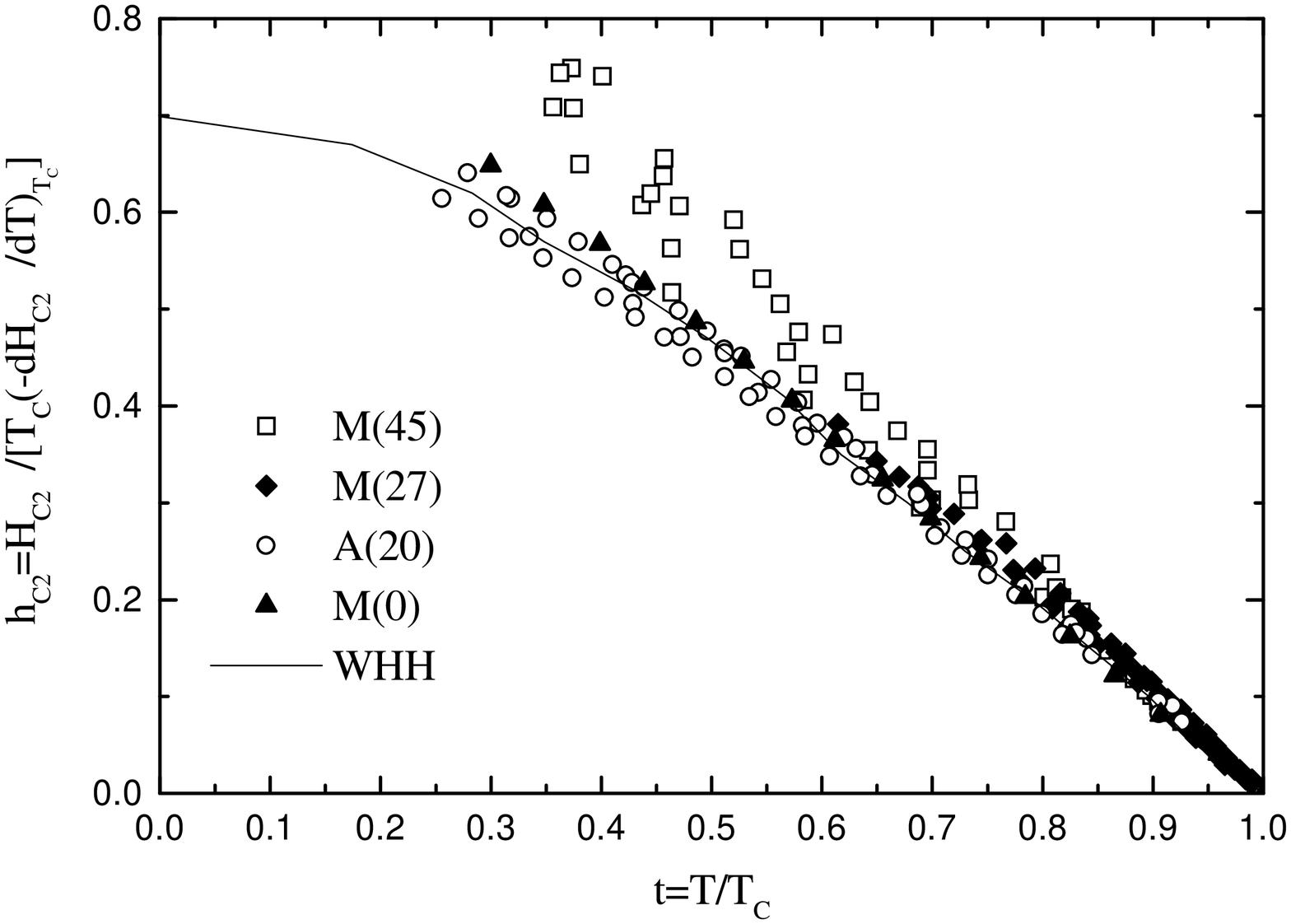}
\end{center}
\caption{\label{fig2}{Reduced perpendicular magnetic field
$h_{c2}$ versus reduced temperature $t$ for all the analyzed
samples. Diamonds refer to the samples of the series M(45); open
squares to the samples of the series M(27); circles to the
samples of the series A(20) and full triangles to the sample
M(0). The solid line is the WHH theoretical curve obtained for
$\alpha=\lambda_{so}=0$.}}
\end{figure}

The upper critical field measurement allows us to investigate the
nature of the pair-breaking mechanism present in our
superconducting-spin glass multilayers. Figure \ref{fig2} shows
the reduced perpendicular critical magnetic field $h_{c2}=H_{c2
\perp}/[T_c(-dH_{c2 \perp}/dT)|_{T=T_c}]$ as a function of the
reduced temperature $t=T/T_c$ for all the investigated samples.
The data have been analyzed in the framework of the WHH theory
which widely describes the $H_{c2}(T)$ behavior of bulk type-II
superconductors, including the case where the effect of applied
magnetic field on the electron spin magnetic moments cannot be
neglected. In particular Pauli spin paramagnetism and spin-orbit
scattering are taken into account, respectively through the
parameters $\alpha$ and $\lambda_{so}$, which appear in the
implicit equation for the reduced field $h_{c2}$\cite{WHH}:

\begin{equation}
ln\Bigg({1 \over t}\Bigg)=\Bigg({1 \over 2} + {i \lambda_{so}
\over 2 \gamma}\Bigg) \Psi\Bigg({1 \over 2} + {\bar h_{c2} + {1
\over 2}\lambda_{so} + i \gamma \over 2 t}\Bigg)+ \Bigg({1 \over
2} - {i \lambda_{so} \over 2 \gamma}\Bigg) \Psi\Bigg({1 \over 2}
+ {\bar h_{c2} + {1 \over 2}\lambda_{so} - i \gamma \over 2
t}\Bigg)-\Psi\Bigg({1 \over 2}\Bigg)
\end{equation}

\noindent where $\psi$ is the digamma function, $\bar h_{c2}=(4/
\pi^2)h_{c2}$ and $\gamma=[(\alpha \bar
h_{c2})^2-((1/2)\lambda{so})^2]^{1/2}$.

While data for all the multilayers with high Mn percentage are
not described by the WHH theory,  for the samples with Mn
percentage up to 2.7 all the experimental points collapse on the
WHH curve calculated for the case $\alpha=\lambda_{so}=0$. This
quite surprising result indicates that a small percentage of Mn
does not significantly influence the $H_{c2 \perp}(T)$ curves and
Nb/CuMn multilayers behave, at least for temperatures down to
$t=0.3$, like ordinary type-II superconductors. $H_{c2 \perp}(T)$
measurements are then less sensitive to Mn concentration than
measurements of $T_c$ versus $d_M$. A nonmonotonic $T_c(d_M)$
dependence was observed even for 0.7 \% of Mn\cite{Mercaldo},
revealing an unconventional proximity effect in the system, while
2.7 \% of Mn is still not sufficient to cause an appreciable
deviation from a conventional $H_{c2 \perp}(T)$ behavior.

In figure \ref{fig2} measurements for a Nb/Cu multilayer, the
sample M(0), are also shown. Again these data collapse on the WHH
curve with $\alpha=\lambda_{so}=0$. Same results have been
obtained for Nb/Pd\cite{Cocco} multilayers. On the other hand
data from samples with high Mn percentage cannot be fitted to the
WHH theory even in the case $\alpha \neq 0$ and $\lambda_{so} \neq
0$. In fact in the $h_{c2}-t$ plane all these data lie above the
points obtained in the small Mn percentage case, while
theoretical curves with $\alpha$, $\lambda_{so} \neq 0$ are
always below the $\alpha = \lambda_{so} = 0$ curve. Similar
results apply for V/Fe\cite{Korevaar} and Nb/Gd\cite{Strunk}
multilayers and Nb/Pd$_{1-x}$Fe$_x$/Nb triple
layers\cite{Vonlohn}, with $x \neq 0$, when plotted in the WHH
fashion. Also in these cases the data lie above the WHH curve
with $\alpha=\lambda_{so}=0$. These results show that an
additional pair breaking mechanism, which is not taken into
account in the WHH theory, is present both in
superconducting/ferromagnetic and some superconducting/spin glass
multilayers, such as Nb/CuMn with high Mn percentage ($>$ 2.7
\%). In this case a realistic explanation requires theories which
explicitly take into account the magnetic nature of the non
superconducting material in the multilayers\cite{Radovic2}.

\section{CONCLUSIONS}

Measurements of $H_{c2 \perp}(T)$ have been performed on several
Nb/CuMn multilayers with different Mn percentage in the CuMn
layers, also in the presence of regular array of antidots. It is
found that the $H_{c2 \perp}(T)$ curves for samples having low Mn
percentage are described by a conventional theory for type-II
superconductors despite the magnetic nature of the samples,
regardless of the layering and of a more complicated structure,
i.e. if a regular array of antidots is present.

\vspace{10.in} Table I. Relevant sample parameters. See the text
for the meaning of the listed quantities.

\begin{tabular}{|c||c|c|c|c|c|c|c|}
\hline\hline

\rm Sample &
$d_{S}$ $(\rm \AA)$
& $d_{M}$ $(\rm \AA)$
& \% Mn
& $\rm N_{bil}$
& $T_c(K)$
& $\zeta$
&  \rm Antidot lattice \\
\hline
M(27)1 & 260 & 6 & 2.7 & 10 & 6.02 & 1.0 & No \\
M(27)2 & 260 & 9 & 2.7 & 10 & 5.42 &  1.0 & No \\
M(27)3 & 260 & 11 & 2.7 & 10 & 4.96 & 1.2 & No \\
M(27)4 & 260 & 16 & 2.7 & 10 & 4.22 & 1.8 & No \\
M(27)5 & 260 & 19 & 2.7 & 10 & 4.03 & 3.3 & No \\
M(27)6 & 260 & 24 & 2.7 & 10 & 4.58 & 3.1 & No \\
M(27)7 & 260 & 29 & 2.7 & 10 & 4.50 & 5.0 & No \\
M(45)1 & 350 & 4 & 4.5 & 10  & 6.67 & 1.37 & No \\
M(45)2 & 350 & 11 & 4.5 & 10 & 5.46 & 1.74 & No \\
M(45)3 & 350 & 15 & 4.5 & 10 & 3.78 & 1.45 & No \\
M(45)4 & 350 & 29 & 4.5 & 10 & 3.67 & 7.23 & No \\
M(45)5 & 350 & 32 & 4.5 & 10 & 3.61 & 5.41 & No \\
A(20)1 & 250 & 8  & 2 & 6 & 7.54 & 1.46 & Yes \\
A(20)2 & 250 & 12 & 2 & 6 & 7.38 & 1.36 & Yes \\
A(20)3 & 250 & 20 & 2 & 6 & 6.96 & 1.41 & Yes \\
A(20)4 & 250 & 24 & 2 & 6 & 6.66 & 1.72 & Yes \\
A(20)5 & 250 & 28 & 2 & 6 & 6.5 & 1.76 & Yes \\
M(0)& 200 & 200 & 0 & 10 & 6.67 & 1.6 & No \\
\hline\hline
\end{tabular}






\vskip1cm

\begin{references}

\bibitem{Degennes} P. G. de Gennes and E. Guyon, Phys. Lett. {\bf 3}, 168 (1963).

\bibitem{Hauser1}  J. J. Hauser, H. C. Theuerer, and N. R. Werthamer, Phys. Rev. {\bf 136}, A637 (1964).

\bibitem{Hauser2}  J. J. Hauser, H. C. Theuerer, and N. R. Werthamer, Phys. Rev. {\bf 142}, 118 (1966).

\bibitem{Jin} B. Y. Jin and J. B. Ketterson, Adv. Phys. {\bf 38}, 189 (1989).

\bibitem{Radovic1} Z. Radovi\'c, M. Ledvij, L. Dobrosavlijevi\'c-Gruij\'c, A. I. Buzdin, and J. R. Clem, Phys. Rev. B {\bf 44}, 759 (1991).

\bibitem{Korevaar} P. Koorevaar, Y. Suzuki, R. Coehoorn, and J. Aarts, Phys. Rev. B {\bf 49}, 441 (1994).

\bibitem{Verbank} G. Verbanck, C. D. Potter, R. Schad, P. Belien, V. V. Moshchalkov, and Y. Bruynseraede,  Physica C {\bf 235-240}, 3295 (1994).

\bibitem{Strunk} C. Strunk, C. S\"urgers, U. Paschen, and H.v. L\"ohneysen, Phys. Rev. B {\bf 49}, 4053 (1994).

\bibitem{Jiang} J. S. Jiang, D. Davidovi\'c, D. H. Reich, and C. L. Chien, Phys. Rev. Lett. {\bf 74}, 314 (1995).

\bibitem{Mughe} Th. M\"uhge, N. N. Garif'yanov, Yu. V. Goryunov, G. G. Khaliullin, L. R. Tagirov, K. Westrholt, I. A. Garifullin, and H. Zabel, Phys. Rev. Lett. {\bf 77}, 1857 (1996).

\bibitem{Mercaldo}  L. V. Mercaldo, C. Attanasio, C. Coccorese, L,
Maritato, S. L. Prischepa, and M. Salvato, Phys. Rev. B {\bf 53},
14040 (1996).

\bibitem{Attanasio} C. Attanasio, C. Coccorese, L. V. Mercaldo, S. L. Prischepa, M. Salvato, and L. Maritato, Phys. Rev. B {\bf 57}, 14411 (1998).

\bibitem{Radovic2} Z. Radovi\'c, L. Dobrosavlijevi\'c-Gruij\'c, A. I. Buzdin, and J. R. Clem, Phys. Rev. B {\bf 38}, 2388 (1988).

\bibitem{PhysicaC1}  C. Attanasio, C. Coccorese, L. V. Mercaldo, M. Salvato, L. Maritato, S. L. Prischepa, C. Giannini, C. Tapfer, L. Ortega, and
F. Comin , Physica C {\bf 312}, 112 (1999).

\bibitem{Jap}  C. Attanasio,  L. Maritato, M. Salvato, S. L. Prischepa,
B. N. Engel, and C. M. Falco, J. Appl. Phys. {\bf 77}, 2081 (1995).

\bibitem{Philmag} C. Attanasio, T. Di Luccio, L. V. Mercaldo, S. L. Prischepa, R. Russo, M. Salvato, L. Maritato, and S. Barbanera, Philosophical Magazine {\bf 80}, 875 (2000).

\bibitem{Prbsubmitted} C. Attanasio, T. Di Luccio, L.V. Mercaldo, S.L. Prischepa, R. Russo, M. Salvato, L. Maritato, S. Barbanera, and A. Tuissi, Phys. Rev. B accepted  for publication.

\bibitem{WHH} N. R. Werthamer, E. Helfand, and P. C. Honenberg, Phys. Rev. {\bf 147}, 295 (1966).

\bibitem{Larkin} A. I. Larkin and Yu. N. Ovchinnikov, in {\sl Non Equilibrium Superconductivity}, edited by P. N. Langenberg and A. I. Larkin (North Holland, Amsterdam, 1986).

\bibitem{Cocco}  C. Coccorese, C. Attanasio, L. V. Mercaldo, M. Salvato, L. Maritato, J. M. Slaughter, C. M. Falco, S. L. Prischepa, and B. I. Ivlev,
Phys. Rev. B {\bf 575}, 7922 (1998).

\bibitem{Vonlohn} M. Sch\"ock, C. S\"urgers, and H.v. L\"ohneysen, Eur.Phys. J. B {\bf 14}, 1 (2000).


\end{references}
\end{document}